\documentclass[9pt,twocolumn,twoside,]{pinp}

\providecommand{\tightlist}{%
  \setlength{\itemsep}{0pt}\setlength{\parskip}{0pt}}

\usepackage[T1]{fontenc}
\usepackage[utf8]{inputenc}

\definecolor{pinpblue}{HTML}{185FAF}  
\definecolor{pnasbluetext}{RGB}{0,101,165} 

\title{Redis for Market Monitoring}

\author[1]{Dirk Eddelbuettel}

  \affil[1]{Department of Statistics, University of Illinois,
Urbana-Champaign, IL, USA}

\leadauthor{Eddelbuettel}

\begin{abstract}
This note shows how to use Redis cache (near-)real-time market data, and
utilise its publish/subscribe (``pub/sub'') facility to distribute the
data.
\end{abstract}

\dates{This version was compiled on \today}

\doifooter{\url{https://cran.r-project.org/package=RcppRedis}}

\pinpfootercontents{Redis Market Monitoring}

\begin{document}

\verticaladjustment{-2pt}

\maketitle
\thispagestyle{firststyle}
\ifthenelse{\boolean{shortarticle}}{\ifthenelse{\boolean{singlecolumn}}{\abscontentformatted}{\abscontent}}{}


\hypertarget{overview}{%
\section{Overview}\label{overview}}

\href{https://redis.io}{Redis} \citep{Redis} is a popular, powerful, and
widely-used `in-memory database-structure store' or server. We provide a
brief introduction to it in a sibbling vignette
\citep{Eddelbuettel:2022:Redis} that is also included in package
\textbf{RcppRedis} \citep{CRAN:RcppRedis}.

This note describes an interesting use case and illustrates both the
ability of \textsf{Redis} to act as a (short-term) data cache (for which
\href{https://redis.io}{Redis} is very frequently used) but also rely on
its ability to act as ``pub/sub'' message broker. The ``pub/sub'' (short
for ``publish/subscribe'') framework is common to distribute data in a
context where (possibly a large number of) ``subscribers'' consume data
provided by one or a few services, often on a local network. Entire
libraries and application frameworks such as
\href{https://zeromq.org/}{ZeroMQ} by \cite{ZeroMQ} (and literally
hundreds more) have pub/sub at its core. But as this note shows, one may
not need anything apart from a (possibly already existing)
\href{https://redis.io}{Redis} client.

\hypertarget{use-case-market-data}{%
\section{Use Case: Market Data}\label{use-case-market-data}}

\hypertarget{basics}{%
\subsection{Basics}\label{basics}}

Monitoring financial market data is a very common task, and many
applications address it. In package \textbf{dang} we provide a function
\texttt{intradayMarketMonitor()} which extends earlier work by
\cite{Ulrich:2021:gist} and does just that for the SP500 index and its
symbol \^{}GSPC (at Yahoo! Finance). For non-tradeable index symbols
such as \^{}GSPC one can retrieve near-``real-time'' updates which is
nice. We put ``real-time'' in quotes here as there are of course delays
in the transmission from the exchange or index provide to a service such
as Yahoo! and then down a retail broadband line to a consumer. Yet it is
``close'' to real-time---as opposed to explicitly delayed data that we
cover below. So \texttt{intradayMarketMonitor()} runs in an endless
loop, updates the symbol and plot, and after market close once writes
its history into an RDS file so that a restart can access some history.
It is nicely minimal and self-contained design.

\begin{figure}[htb]
  \begin{center}
    \includegraphics[width=3.5in,height=4.75in]{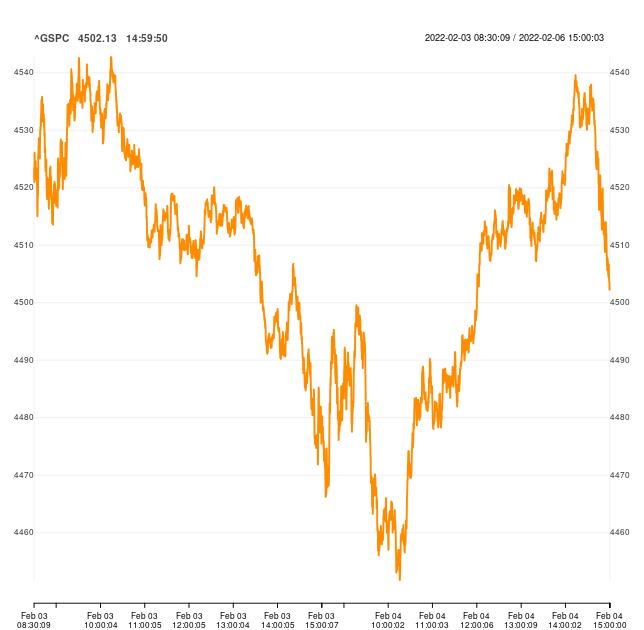}
    \caption{Intraday Market Monitoring Example}
    \label{fig:intraday-market-monitoring}
  \end{center}
\end{figure}

Figure \ref{fig:intraday-market-monitoring} shows a plot resulting from
calling the function on a symbol, here again \^{}GSPC, when two days of
history have been accumulated. (The plot was generated on a weekend with
the preceding Friday close providing the last data point.)

\hypertarget{possible-shortcomings}{%
\subsection{Possible Shortcomings}\label{possible-shortcomings}}

Some of the short-comings of the approach in
\texttt{intradayMarketMonitor()} and \cite{Ulrich:2021:gist} are

\begin{itemize}
\tightlist
\item
  use of one R process per symbol
\item
  same process used for monitoring and plotting
\item
  no persistence until end of day
\end{itemize}

Moreover, the `real-time' symbol for the main market index is available
only during (New York Stock Exchange) market hours. Yet sometimes one
wants to gauge a market reaction or `mood' at off-market hours.

So with this, the idea arose to decouple market data \emph{acquisition
and caching} from actual \emph{visualization} or other monitoring. This
would also permit distributing the tasks over several machines: for
example an `always-on' monitoring machine could always track the data
and store it for other `on-demand' machines or applications to access
it. And as we have seen, \href{https://redis.io}{Redis} makes for a fine
data `caching' mechanism.

\hypertarget{building-a-market-monitor}{%
\section{Building A Market Monitor}\label{building-a-market-monitor}}

\hypertarget{data}{%
\subsection{Data}\label{data}}

The \textbf{quantmod} package by \cite{CRAN:quantmod} provides a
function \texttt{getQuote()} we can use to obtain data snapshots. We
will look at \^{}GSPC as before but also ES=F, the Yahoo! Finance symbol
for the `rolling front contract' for the SP500 Futures trading at CME
Globex under symbol ES. (We will not get into details on futures
contracts here as the topic is extensively covered elsewhere. We will
just add that equity futures tend to trade in only one contract (``no
curve'') and roll to the next quarterly expiration at particular dates
well established and known by market practice.)

\begin{Shaded}
\begin{Highlighting}[]
\FunctionTok{suppressMessages}\NormalTok{(}\FunctionTok{library}\NormalTok{(quantmod))}
\NormalTok{res }\OtherTok{\textless{}{-}} \FunctionTok{getQuote}\NormalTok{(}\FunctionTok{c}\NormalTok{(}\StringTok{"\^{}GSPC"}\NormalTok{, }\StringTok{"ES=F"}\NormalTok{, }\StringTok{"SPY"}\NormalTok{))}
\NormalTok{res[,}\DecValTok{1}\SpecialCharTok{:}\DecValTok{3}\NormalTok{] }\CommentTok{\# omitting chg, OHL, Vol}
\CommentTok{\#                 Trade Time    Last   Change}
\CommentTok{\#  \^{}GSPC 2022{-}03{-}15 17:28:36 4262.45 89.34033}
\CommentTok{\#  ES=F  2022{-}03{-}15 19:48:58 4261.75 {-}0.25000}
\CommentTok{\#  SPY   2022{-}03{-}15 16:00:01  426.17  9.17001}
\end{Highlighting}
\end{Shaded}

The preceding code display shows how the \textbf{quantmod}
\citep{CRAN:quantmod} funtion \texttt{getQuote()} can access index data
(symbol `\^{}GSPC'), futures data (symbol `ES=F' as the rolling front
contract) as well as equity / ETF data (symbol `SPY').

\hypertarget{storing-and-publishing}{%
\subsection{Storing and Publishing}\label{storing-and-publishing}}

Given per-security rows of data as shown in the preceding example, we
can use \href{https://redis.io}{Redis} to store the data using the
timestamp as a sorting criterion in a per-symbol stack. The `sorted set'
data structure is very appropriate for this. The function
\texttt{get\_data()} transforms the result of \texttt{getQuote()} into a
named numeric vector suitable for our use of `sorted sets'.

\begin{Shaded}
\begin{Highlighting}[]
\NormalTok{get\_data }\OtherTok{\textless{}{-}} \ControlFlowTok{function}\NormalTok{(symbol) \{}
\NormalTok{    quote }\OtherTok{\textless{}{-}} \FunctionTok{getQuote}\NormalTok{(symbol)}
\NormalTok{    vec }\OtherTok{\textless{}{-}} \FunctionTok{c}\NormalTok{(}\AttributeTok{Time =} \FunctionTok{as.numeric}\NormalTok{(quote}\SpecialCharTok{$}\StringTok{\textasciigrave{}}\AttributeTok{Trade Time}\StringTok{\textasciigrave{}}\NormalTok{),}
             \AttributeTok{Close =}\NormalTok{ quote}\SpecialCharTok{$}\NormalTok{Last,}
             \AttributeTok{Change =}\NormalTok{ quote}\SpecialCharTok{$}\NormalTok{Change,}
             \AttributeTok{PctChange =}\NormalTok{ quote}\SpecialCharTok{$}\StringTok{\textasciigrave{}}\AttributeTok{\% Change}\StringTok{\textasciigrave{}}\NormalTok{,}
             \AttributeTok{Volume =}\NormalTok{ quote}\SpecialCharTok{$}\NormalTok{Volume)}
\NormalTok{    vec}
\NormalTok{\}}
\end{Highlighting}
\end{Shaded}

Similarly, given the symbol, we can also `publish' a datum with the
current values and timestamp. In the example application included with
\textbf{Redis}, this is done by relying on the following short function
which receives the current data record and then stores and publish it.

\begin{Shaded}
\begin{Highlighting}[]
\NormalTok{store\_data }\OtherTok{\textless{}{-}} \ControlFlowTok{function}\NormalTok{(vec, symbol) \{}
\NormalTok{    redis}\SpecialCharTok{$}\FunctionTok{zadd}\NormalTok{(symbol, }\FunctionTok{matrix}\NormalTok{(vec, }\DecValTok{1}\NormalTok{))}
\NormalTok{    redis}\SpecialCharTok{$}\FunctionTok{publish}\NormalTok{(symbol, }\FunctionTok{paste}\NormalTok{(vec,}\AttributeTok{collapse=}\StringTok{";"}\NormalTok{))}
\NormalTok{\}}
\end{Highlighting}
\end{Shaded}

In this example, the \texttt{redis} instance is a script-level global
symbol. This could easily be rewritten where it is also be passed into
the function, and \texttt{vec} is a simple vector of observations
procured by \texttt{getQuote()} as discussed in the preceding code
example. The timestamp is transformed into a numeric value making the
vector all-numeric which the format used by \texttt{zadd()} to added a
`sorted' (by the timestamp) numeric one-row matrix. Beside storing the
data, we also publish it via \textsf{Redis} on channel named as the
symbol. Here the numeric data is simply concatenated with a \texttt{;}
as separator and sent as text.

The core functionality in the main loop is then as follows below where
we also omitted some of the error or status messaging for brevity.

In that example, the change is volume is used as a `tell' for actual new
data. This works reliably for the (main futures) markets we follow here
which have essentially constant trading activity. When some tranquil
periods occur, the gaps between stored and published data points may be
longer than the default sleep period of ten seconds used here.

\begin{Shaded}
\begin{Highlighting}[]
\NormalTok{    y }\OtherTok{\textless{}{-}} \FunctionTok{try}\NormalTok{(}\FunctionTok{get\_data}\NormalTok{(symbol), }\AttributeTok{silent =} \ConstantTok{TRUE}\NormalTok{)}
    \ControlFlowTok{if}\NormalTok{ (}\FunctionTok{inherits}\NormalTok{(y, }\StringTok{"try{-}error"}\NormalTok{)) \{}
        \FunctionTok{msg}\NormalTok{(curr\_t, }\StringTok{"Error ..."}\NormalTok{) }\CommentTok{\# rest omitted}
\NormalTok{        errored }\OtherTok{\textless{}{-}} \ConstantTok{TRUE}
        \FunctionTok{Sys.sleep}\NormalTok{(}\DecValTok{15}\NormalTok{)}
        \ControlFlowTok{next}
\NormalTok{    \} }\ControlFlowTok{else} \ControlFlowTok{if}\NormalTok{ (errored) \{}
\NormalTok{        errored }\OtherTok{\textless{}{-}} \ConstantTok{FALSE}
        \FunctionTok{msg}\NormalTok{(curr\_t, }\StringTok{"...recovered"}\NormalTok{)}
\NormalTok{    \}}
\NormalTok{    v }\OtherTok{\textless{}{-}}\NormalTok{ y[}\StringTok{"Volume"}\NormalTok{]}
    \ControlFlowTok{if}\NormalTok{ (v }\SpecialCharTok{!=}\NormalTok{ prevVol) \{}
        \FunctionTok{store\_data}\NormalTok{(y, symbol)}
        \FunctionTok{msg}\NormalTok{(curr\_t, }\StringTok{"Storing ..."}\NormalTok{) }\CommentTok{\# same}
\NormalTok{    \}}
\NormalTok{    prevVol }\OtherTok{\textless{}{-}}\NormalTok{ v}
    \FunctionTok{Sys.sleep}\NormalTok{(}\DecValTok{10}\NormalTok{)}
\end{Highlighting}
\end{Shaded}

The remainder of the `acquiring data and storing in \textsf{Redis}' code
is similar to the non-\textsf{Redis} using variant
\texttt{intradayMarketMonitor()} in \textbf{dang} \citep{CRAN:dang} that
is based on the earlier work by \cite{Ulrich:2021:gist}.

\hypertarget{retrieving-and-subscribing}{%
\subsection{Retrieving and
Subscribing}\label{retrieving-and-subscribing}}

Two core routines to receive data from \textsf{Redis} to plot both read
the most recent stored data at startup, and then grow this data set via
a subscription to the updates published to the channel.

We first show the initial request for all data, which is then subset to
the \(n\) most recent days. We can request `all' data as we also deploy
a helper script referenced in the appendix to keep the overall data
volume that is stored at `manageable' and finite levels. Adding such a
step is important for a process such as this which continually appends
data which, if unchecked, would `eventually' exhaust system resources.

\begin{Shaded}
\begin{Highlighting}[]
\NormalTok{most\_recent\_n\_days }\OtherTok{\textless{}{-}} \ControlFlowTok{function}\NormalTok{(x, }\AttributeTok{n=}\DecValTok{2}\NormalTok{,}
                               \AttributeTok{minobs=}\DecValTok{1500}\NormalTok{) \{}
\NormalTok{    tt }\OtherTok{\textless{}{-}} \FunctionTok{table}\NormalTok{(}\FunctionTok{as.Date}\NormalTok{(}\FunctionTok{index}\NormalTok{(x)))}
    \ControlFlowTok{if}\NormalTok{ (}\FunctionTok{length}\NormalTok{(tt) }\SpecialCharTok{\textless{}}\NormalTok{ n) }\FunctionTok{return}\NormalTok{(x)}
\NormalTok{    ht }\OtherTok{\textless{}{-}} \FunctionTok{head}\NormalTok{(}\FunctionTok{tail}\NormalTok{(tt[tt}\SpecialCharTok{\textgreater{}}\NormalTok{minobs], n), }\DecValTok{1}\NormalTok{)}
\NormalTok{    cutoff }\OtherTok{\textless{}{-}} \FunctionTok{paste}\NormalTok{(}\FunctionTok{format}\NormalTok{(}\FunctionTok{as.Date}\NormalTok{(}\FunctionTok{names}\NormalTok{(ht))),}
                    \StringTok{"00:00:00"}\NormalTok{)}
\NormalTok{    newx }\OtherTok{\textless{}{-}}\NormalTok{ x[ }\FunctionTok{index}\NormalTok{(x) }\SpecialCharTok{\textgreater{}=} \FunctionTok{as.POSIXct}\NormalTok{(cutoff) ]}
    \FunctionTok{msg}\NormalTok{(}\FunctionTok{Sys.time}\NormalTok{(), }\StringTok{"most recent data starting at"}\NormalTok{,}
        \FunctionTok{format}\NormalTok{(}\FunctionTok{head}\NormalTok{(}\FunctionTok{index}\NormalTok{(newx),}\DecValTok{1}\NormalTok{)))}
\NormalTok{    newx}
\NormalTok{\}}

\NormalTok{get\_all\_data }\OtherTok{\textless{}{-}} \ControlFlowTok{function}\NormalTok{(symbol, host) \{}
\NormalTok{    m }\OtherTok{\textless{}{-}}\NormalTok{ redis}\SpecialCharTok{$}\FunctionTok{zrange}\NormalTok{(symbol, }\DecValTok{0}\NormalTok{, }\SpecialCharTok{{-}}\DecValTok{1}\NormalTok{)}
    \FunctionTok{colnames}\NormalTok{(m) }\OtherTok{\textless{}{-}} \FunctionTok{c}\NormalTok{(}\StringTok{"Time"}\NormalTok{, }\StringTok{"Close"}\NormalTok{, }\StringTok{"Change"}\NormalTok{,}
                     \StringTok{"PctChange"}\NormalTok{, }\StringTok{"Volume"}\NormalTok{)}
\NormalTok{    y }\OtherTok{\textless{}{-}} \FunctionTok{xts}\NormalTok{(m[,}\SpecialCharTok{{-}}\DecValTok{1}\NormalTok{],}
             \AttributeTok{order.by=}\FunctionTok{anytime}\NormalTok{(}\FunctionTok{as.numeric}\NormalTok{(m[,}\DecValTok{1}\NormalTok{])))}
\NormalTok{    y}
\NormalTok{\}}

\DocumentationTok{\#\# ... some setup}
\NormalTok{x }\OtherTok{\textless{}{-}} \FunctionTok{get\_all\_data}\NormalTok{(symbol, host)}
\NormalTok{x }\OtherTok{\textless{}{-}} \FunctionTok{most\_recent\_n\_days}\NormalTok{(x,ndays)}
\end{Highlighting}
\end{Shaded}

The updates from subscription happen in the main \texttt{while()} loop.
The subscription is set up as follows:

\begin{Shaded}
\begin{Highlighting}[]
\DocumentationTok{\#\# This is the callback func. assigned to a symbol}
\NormalTok{.data2xts }\OtherTok{\textless{}{-}} \ControlFlowTok{function}\NormalTok{(x) \{}
\NormalTok{    m }\OtherTok{\textless{}{-}} \FunctionTok{read.csv}\NormalTok{(}\AttributeTok{text=}\NormalTok{x, }\AttributeTok{sep=}\StringTok{";"}\NormalTok{, }\AttributeTok{header=}\ConstantTok{FALSE}\NormalTok{,}
                  \AttributeTok{col.names=}\FunctionTok{c}\NormalTok{(}\StringTok{"Time"}\NormalTok{, }\StringTok{"Close"}\NormalTok{,}
                              \StringTok{"Change"}\NormalTok{,}\StringTok{"PctChange"}\NormalTok{,}
                              \StringTok{"Volume"}\NormalTok{))}
\NormalTok{    y }\OtherTok{\textless{}{-}} \FunctionTok{xts}\NormalTok{(m[,}\SpecialCharTok{{-}}\DecValTok{1}\NormalTok{,}\AttributeTok{drop=}\ConstantTok{FALSE}\NormalTok{],}
             \FunctionTok{anytime}\NormalTok{(}\FunctionTok{as.numeric}\NormalTok{(m[,}\DecValTok{1}\NormalTok{,}\AttributeTok{drop=}\ConstantTok{FALSE}\NormalTok{])))}
\NormalTok{    y}
\NormalTok{\}}
\CommentTok{\# programmatic version of \textasciigrave{}ES=F\textasciigrave{} \textless{}{-} function(x) ...}
\FunctionTok{assign}\NormalTok{(symbol, .data2xts)}
\NormalTok{redis}\SpecialCharTok{$}\FunctionTok{subscribe}\NormalTok{(symbol)}
\end{Highlighting}
\end{Shaded}

The \texttt{.data2xts()} callback function parses the concatenated
values, and constructs a one-row object \texttt{xts} object. The
\textbf{xts} package by \cite{CRAN:xts} make time-ordered appending of
such data via \texttt{rbind} easy which is what is done in the main
loop:

\begin{Shaded}
\begin{Highlighting}[]
\NormalTok{    y }\OtherTok{\textless{}{-}} \FunctionTok{redisMonitorChannels}\NormalTok{(redis)}
    \ControlFlowTok{if}\NormalTok{ (}\SpecialCharTok{!}\FunctionTok{is.null}\NormalTok{(y)) \{}
\NormalTok{        x }\OtherTok{\textless{}{-}} \FunctionTok{rbind}\NormalTok{(x,y)}
\NormalTok{        x }\OtherTok{\textless{}{-}}\NormalTok{ x[}\SpecialCharTok{!}\FunctionTok{duplicated}\NormalTok{(}\FunctionTok{index}\NormalTok{(x))]}
\NormalTok{    \}}
    \FunctionTok{show\_plot}\NormalTok{(symbol, x)}
\end{Highlighting}
\end{Shaded}

The \texttt{redisMonitorChannels(redis)} is key to our pub/sub mechanism
here. The subscriptions are stored in the \texttt{redis} instance, along
with any optional callbacks. The function will listen to (one or more)
channels using the key \textsf{Redis} function \texttt{listen()} and
consume the next message. The key here is our addition of an optional
per-symbol callback which, if present, is used to process the returned
data. This means that in our application with the \texttt{.data2xts()}
function used as a per-symbol callback, the returned variable \texttt{y}
above is a standard \texttt{xts} object which \texttt{rbind} efficiently
appends to an existing object which is how we grow \texttt{x} here. (For
brevity we have omitted two statements messaging data upgrade process to
the console when running, they are included in the full source file
included in the package.)

\hypertarget{extending-to-multiple-symbol}{%
\section{Extending to Multiple
Symbol}\label{extending-to-multiple-symbol}}

\begin{figure}[htb]
  \begin{center}
    \includegraphics[width=3.5in,height=4.75in]{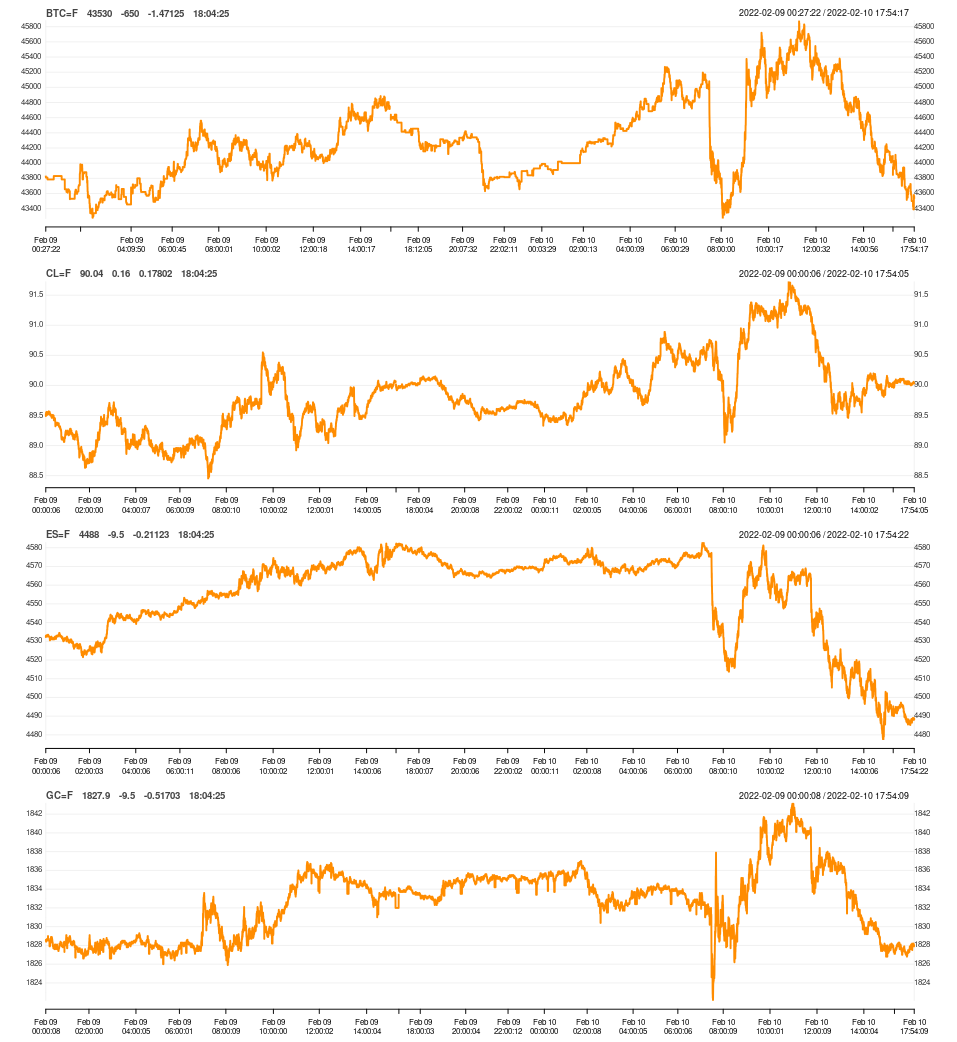}
    \caption{Multi-Symbol Market Monitoring Example}
    \label{fig:multi-symbols-market-monitoring}
  \end{center}
\end{figure}

The pub/sub mechanism is very powerful. Listening to a market symbol,
storing it, and publishing for use on local network enables and
facilitates further use of the data.

Naturally, the idea arises to listen to multiple symbols. At first
glance, one could run one listener process by symbol. The advantage is
the ease of use. A clear disadvantage is the inefficient resource
utilization.

And it turns out that we do not have to. Just how the initial
\texttt{quantmod::getQuote()} call shows access to \emph{several}
symbols at once, we can then process a reply from \texttt{getQuote()}
and store and publish \emph{multiple} symbols on multiple channels. This
is done in files \texttt{intraday-GLOBEX-to-Redis.r} and
\texttt{intraday-GLOBEX-from-Redis.r}. Just like the initial examples
for ES, these files show how to cover several symbols. Here we use for:
Bitcoin, SP500, Gold, and WTI Crude Oil. By sticking to the same
exchanges, here CME Globex, we can use one set of `open' or `close'
rules.

\hypertarget{data-and-publishing}{%
\subsection{Data and Publishing}\label{data-and-publishing}}

The following snippet fetches the data and stores and publishes it.

\begin{Shaded}
\begin{Highlighting}[]
\NormalTok{symbols }\OtherTok{\textless{}{-}} \FunctionTok{c}\NormalTok{(}\StringTok{"BTC=F"}\NormalTok{, }\StringTok{"CL=F"}\NormalTok{, }\StringTok{"ES=F"}\NormalTok{, }\StringTok{"GC=F"}\NormalTok{)}

\NormalTok{get\_data }\OtherTok{\textless{}{-}} \ControlFlowTok{function}\NormalTok{(symbols) \{}
\NormalTok{    quotes }\OtherTok{\textless{}{-}} \FunctionTok{getQuote}\NormalTok{(symbols)}
\NormalTok{    quotes}\SpecialCharTok{$}\NormalTok{Open }\OtherTok{\textless{}{-}}\NormalTok{ quotes}\SpecialCharTok{$}\NormalTok{High }\OtherTok{\textless{}{-}}\NormalTok{ quotes}\SpecialCharTok{$}\NormalTok{Low }\OtherTok{\textless{}{-}}\ConstantTok{NULL}
    \FunctionTok{colnames}\NormalTok{(quotes) }\OtherTok{\textless{}{-}} \FunctionTok{c}\NormalTok{(}\StringTok{"Time"}\NormalTok{,}\StringTok{"Close"}\NormalTok{,}\StringTok{"Change"}\NormalTok{,}
                          \StringTok{"PctChange"}\NormalTok{, }\StringTok{"Volume"}\NormalTok{)}
\NormalTok{    quotes}\SpecialCharTok{$}\NormalTok{Time }\OtherTok{\textless{}{-}} \FunctionTok{as.numeric}\NormalTok{(quotes}\SpecialCharTok{$}\NormalTok{Time)}
\NormalTok{    quotes}
\NormalTok{\}}

\NormalTok{store\_data }\OtherTok{\textless{}{-}} \ControlFlowTok{function}\NormalTok{(res) \{}
\NormalTok{    symbols }\OtherTok{\textless{}{-}} \FunctionTok{rownames}\NormalTok{(res)}
\NormalTok{    res }\OtherTok{\textless{}{-}} \FunctionTok{as.matrix}\NormalTok{(res)}
    \ControlFlowTok{for}\NormalTok{ (symbol }\ControlFlowTok{in}\NormalTok{ symbols) \{}
\NormalTok{        vec }\OtherTok{\textless{}{-}}\NormalTok{ res[symbol,,drop}\OtherTok{=}\ConstantTok{FALSE}\NormalTok{]}
\NormalTok{        redis}\SpecialCharTok{$}\FunctionTok{zadd}\NormalTok{(symbol, vec)}
\NormalTok{        redis}\SpecialCharTok{$}\FunctionTok{publish}\NormalTok{(symbol,}
                      \FunctionTok{paste}\NormalTok{(vec,}\AttributeTok{collapse=}\StringTok{";"}\NormalTok{))}
\NormalTok{    \}}
\NormalTok{\}}
\end{Highlighting}
\end{Shaded}

It is used in the main loop inside a \texttt{try()} statement and error
handler.

\begin{Shaded}
\begin{Highlighting}[]
\NormalTok{    res }\OtherTok{\textless{}{-}} \FunctionTok{try}\NormalTok{(}\FunctionTok{get\_data}\NormalTok{(symbols), }\AttributeTok{silent =} \ConstantTok{TRUE}\NormalTok{)}
    \ControlFlowTok{if}\NormalTok{ (}\FunctionTok{inherits}\NormalTok{(res, }\StringTok{"try{-}error"}\NormalTok{)) \{}
        \FunctionTok{msg}\NormalTok{(curr\_t, }\StringTok{"Error:"}\NormalTok{,}
            \FunctionTok{attr}\NormalTok{(res, }\StringTok{"condition"}\NormalTok{)[[}\StringTok{"message"}\NormalTok{]])}
\NormalTok{        errored }\OtherTok{\textless{}{-}} \ConstantTok{TRUE}
        \FunctionTok{Sys.sleep}\NormalTok{(}\DecValTok{15}\NormalTok{)}
        \ControlFlowTok{next}
\NormalTok{    \} }\ControlFlowTok{else} \ControlFlowTok{if}\NormalTok{ (errored) \{}
\NormalTok{        errored }\OtherTok{\textless{}{-}} \ConstantTok{FALSE}
        \FunctionTok{msg}\NormalTok{(curr\_t, }\StringTok{"...recovered"}\NormalTok{)}
\NormalTok{    \}}
\NormalTok{    v }\OtherTok{\textless{}{-}}\NormalTok{ res[}\DecValTok{3}\NormalTok{, }\StringTok{"Volume"}\NormalTok{]}
    \ControlFlowTok{if}\NormalTok{ (v }\SpecialCharTok{!=}\NormalTok{ prevVol) \{}
        \FunctionTok{store\_data}\NormalTok{(res)}
        \CommentTok{\# msg(...omitted for brevity...)}
\NormalTok{    \}}
\NormalTok{    prevVol }\OtherTok{\textless{}{-}}\NormalTok{ v}
    \FunctionTok{Sys.sleep}\NormalTok{(}\DecValTok{10}\NormalTok{)}
\end{Highlighting}
\end{Shaded}

\hypertarget{retrieving-data}{%
\subsection{Retrieving data}\label{retrieving-data}}

The receiving side of the application works similarly. First, we need to
subscribe to multiple channels:

\begin{Shaded}
\begin{Highlighting}[]
\NormalTok{env }\OtherTok{\textless{}{-}} \FunctionTok{new.env}\NormalTok{() }\CommentTok{\# local environment for callbacks}

\DocumentationTok{\#\# same .data2xts() function  as above}

\DocumentationTok{\#\# With environment \textquotesingle{}env\textquotesingle{}, assign callback}
\DocumentationTok{\#\# function for each symbol}
\NormalTok{res }\OtherTok{\textless{}{-}} \FunctionTok{sapply}\NormalTok{(symbols, }\ControlFlowTok{function}\NormalTok{(symbol) \{}
    \DocumentationTok{\#\# progr. version of \textasciigrave{}ES=F\textasciigrave{} \textless{}{-} function(x) ...}
    \FunctionTok{assign}\NormalTok{(symbol, .data2xts, }\AttributeTok{envir=}\NormalTok{env)}
\NormalTok{    redis}\SpecialCharTok{$}\FunctionTok{subscribe}\NormalTok{(symbol)}
\NormalTok{\})}
\end{Highlighting}
\end{Shaded}

We then use a slighly generalized listener:

\begin{Shaded}
\begin{Highlighting}[]
\DocumentationTok{\#\# Callback handler for convenience}
\NormalTok{multiSymbolRedisMonitorChannels }\OtherTok{\textless{}{-}}
    \ControlFlowTok{function}\NormalTok{(context,}
             \AttributeTok{type=}\StringTok{"rdata"}\NormalTok{, }\AttributeTok{env=}\NormalTok{.GlobalEnv) \{}
\NormalTok{    res }\OtherTok{\textless{}{-}}\NormalTok{ context}\SpecialCharTok{$}\FunctionTok{listen}\NormalTok{(type)}
    \ControlFlowTok{if}\NormalTok{ (}\FunctionTok{length}\NormalTok{(res) }\SpecialCharTok{!=} \DecValTok{3} \SpecialCharTok{||}
\NormalTok{        res[[}\DecValTok{1}\NormalTok{]] }\SpecialCharTok{!=} \StringTok{"message"}\NormalTok{) }\FunctionTok{return}\NormalTok{(res)}
    \ControlFlowTok{if}\NormalTok{ (}\FunctionTok{exists}\NormalTok{(res[[}\DecValTok{2}\NormalTok{]], }\AttributeTok{mode=}\StringTok{"function"}\NormalTok{,}
               \AttributeTok{envir=}\NormalTok{env)) \{}
\NormalTok{        data }\OtherTok{\textless{}{-}} \FunctionTok{do.call}\NormalTok{(res[[}\DecValTok{2}\NormalTok{]],}
                        \FunctionTok{as.list}\NormalTok{(res[[}\DecValTok{3}\NormalTok{]]),}
                        \AttributeTok{envir=}\NormalTok{env)}
\NormalTok{        val }\OtherTok{\textless{}{-}} \FunctionTok{list}\NormalTok{(}\AttributeTok{symbol=}\NormalTok{res[[}\DecValTok{2}\NormalTok{]],}
                    \AttributeTok{data=}\NormalTok{data)}
        \FunctionTok{return}\NormalTok{(val)}
\NormalTok{    \}}
\NormalTok{    res}
\NormalTok{\}}
\end{Highlighting}
\end{Shaded}

The \texttt{listen} methods returns an object which is checked for
correct length and first component. If appropriate, the second element
is the channel symbol so if a callback function of the same names
exists, it is called with the third element, the `payload'. This creates
the familiar \texttt{xts} object with is return along with the symbol in
a two-element list.

The data is consumed in the \texttt{while} loop in a very similar
fashion to the one-symbol case, but we now unpack the loop and operate
on the appropriate data element.

\begin{Shaded}
\begin{Highlighting}[]
    \DocumentationTok{\#\# monitor channels in context of \textquotesingle{}env\textquotesingle{}}
\NormalTok{    rl }\OtherTok{\textless{}{-}} \FunctionTok{multiSymbolRedisMonitorChannels}\NormalTok{(redis,}
                                          \AttributeTok{env=}\NormalTok{env)}
    \ControlFlowTok{if}\NormalTok{ (}\FunctionTok{is.list}\NormalTok{(rl)) \{}
\NormalTok{        sym }\OtherTok{\textless{}{-}}\NormalTok{ rl[[}\StringTok{"symbol"}\NormalTok{]]}
\NormalTok{        x[[sym]] }\OtherTok{\textless{}{-}} \FunctionTok{rbind}\NormalTok{(x[[sym]], rl[[}\StringTok{"data"}\NormalTok{]])}
\NormalTok{        z }\OtherTok{\textless{}{-}} \FunctionTok{tail}\NormalTok{(x[[sym]],}\DecValTok{1}\NormalTok{)}
        \ControlFlowTok{if}\NormalTok{ (sym }\SpecialCharTok{==}\NormalTok{ symbols[}\DecValTok{3}\NormalTok{]) }\FunctionTok{msg}\NormalTok{(}\CommentTok{\#...omitted...)}
    \ErrorTok{\}} \ControlFlowTok{else}\NormalTok{ \{}
        \FunctionTok{msg}\NormalTok{(}\FunctionTok{index}\NormalTok{(now\_t), }\StringTok{"null data in y"}\NormalTok{)}
\NormalTok{    \}}
    \FunctionTok{show\_plot}\NormalTok{(symbols, x)}
\end{Highlighting}
\end{Shaded}

Finally, the plot function simply plots for all symbols in the
\texttt{symbols} vector.

Overall, this setup is robust to data `surprises' as the \texttt{try()}
mechanism implements an error recovery in cases of temporary network or
remote server issues. The overall design is simple: each of the two
files for, respectively, receiving-and-storing data and
accessing-and-visualizing, contains only a few short helper functions
(most of which where shown above) and a core \texttt{while()} loop. We
have had these running uninterrupted and without issues for months on
end.

\hypertarget{summary}{%
\section{Summary}\label{summary}}

We describe a simple yet efficient mechanism to capture and publish
`live' market data by relying on \textsf{Redis} via the
\textbf{RcppRedis} package.

\hypertarget{acknowledgements}{%
\section{Acknowledgements}\label{acknowledgements}}

Joshua Ulrich provided a first useable monotoring loop for a life symbol
which is gratefully acknowledged, as are numerous discussions about
\textbf{quantmod} and other packages. Bryan Lewis not only put an
elegant and working pub/sub mechanism in his \textbf{rredis}, but also
ported it into a very elegant callback-based solution in package
\textbf{RcppRedis}. These features, and this monitoring application,
would not exists without the help of either Josh or Bryan.

\hypertarget{appendix}{%
\section{Appendix}\label{appendix}}

\hypertarget{data-growth}{%
\subsection{Data Growth}\label{data-growth}}

The scripts do not write the data to \textsf{Redis} with a
`time-to-live' (TTL) expiry. This means the database is growing. A
simple way to limit the growth is to invoke a pruning script from
\texttt{cron} once a week. We include a simple script in the
\texttt{pub-sub/} directory of the package.


\bibliography{redis}

\begin{thebibliography}{8}
\newcommand{\enquote}[1]{``#1''}
\providecommand{\natexlab}[1]{#1}
\providecommand{\url}[1]{\texttt{#1}}
\providecommand{\urlprefix}{URL }
\expandafter\ifx\csname urlstyle\endcsname\relax
  \providecommand{\doi}[1]{doi:\discretionary{}{}{}#1}\else
  \providecommand{\doi}{doi:\discretionary{}{}{}\begingroup
  \urlstyle{rm}\Url}\fi
\providecommand{\eprint}[2][]{\url{#2}}

\bibitem[{Eddelbuettel(2021)}]{CRAN:dang}
Eddelbuettel D (2021).
\newblock \emph{dang: 'Dang' Associated New Goodies}.
\newblock R package version 0.0.15,
  \urlprefix\url{https://CRAN.R-project.org/package=dang}.

\bibitem[{Eddelbuettel(2022)}]{Eddelbuettel:2022:Redis}
Eddelbuettel D (2022).
\newblock \enquote{{A Brief Introduction to Redis}.}
\newblock \doi{10.48550/arXiv.2203.06559}.

\bibitem[{Eddelbuettel and Lewis(2022)}]{CRAN:RcppRedis}
Eddelbuettel D, Lewis BW (2022).
\newblock \emph{RcppRedis: 'Rcpp' Bindings for 'Redis' using the 'hiredis'
  Library}.
\newblock R package version 0.2.0,
  \urlprefix\url{https://CRAN.R-Project.org/package=RcppRedis}.

\bibitem[{Hintjens and Sustrik(2010)}]{ZeroMQ}
Hintjens P, Sustrik M (2010).
\newblock \enquote{ZeroMQ: An open-source universal messaging library.}
\newblock \url{https://zeromq.org}.

\bibitem[{Ryan and Ulrich(2020{\natexlab{a}})}]{CRAN:quantmod}
Ryan JA, Ulrich JM (2020{\natexlab{a}}).
\newblock \emph{quantmod: Quantitative Financial Modelling Framework}.
\newblock R package version 0.4.18,
  \urlprefix\url{https://CRAN.R-project.org/package=quantmod}.

\bibitem[{Ryan and Ulrich(2020{\natexlab{b}})}]{CRAN:xts}
Ryan JA, Ulrich JM (2020{\natexlab{b}}).
\newblock \emph{xts: eXtensible Time Series}.
\newblock R package version 0.12.1,
  \urlprefix\url{https://CRAN.R-project.org/package=xts}.

\bibitem[{Sanfilippo(2009)}]{Redis}
Sanfilippo S (2009).
\newblock \enquote{Redis In-memory Data Structure Server.}
\newblock \url{https://redis.io}.

\bibitem[{Ulrich(2021)}]{Ulrich:2021:gist}
Ulrich JM (2021).
\newblock \enquote{Market-Monitoring with R.}
\newblock
  \url{https://gist.github.com/joshuaulrich/ee11ef67b1461df399b84efd3c8f9f67#file-intraday-sp500-r}.

\end{thebibliography}
\bibliographystyle{jss}

\end{document}